\documentstyle[12pt,epsf]{article}

\def\inseps#1#2{\def\epsfsize##1##2{#2##1} \centerline{\epsfbox{#1}}}

\begin{document}

\setlength{\baselineskip}{7mm}

\title{\bf Monte Carlo simulation and global optimization without parameters}
\author{Bobby Hesselbo and R. B. Stinchcombe \\ Theoretical Physics, University
of Oxford,\\
1 Keble Road, Oxford, OX1 3NP, United Kingdom.}

\maketitle

\begin{abstract}
We propose a new ensemble for Monte Carlo simulations, in which each
state is assigned a statistical weight $1/k$, where $k$ is the number
of states with smaller or equal energy. This ensemble has robust
ergodicity properties and gives significant weight to the ground
state, making it effective for hard optimization problems. It can be
used to find free energies at all temperatures and picks up aspects of
critical behaviour (if present) without any parameter tuning. We test
it on the travelling salesperson problem, the Edwards-Anderson spin
glass and the triangular antiferromagnet.
\end{abstract}


\vfill
\newpage

The method of Monte Carlo simulation has proved very useful for
studying the thermodynamic properties of model systems with moderately
many degrees of freedom. The idea is to sample the system's phase
space stochastically, using a computer to generate a series of random
configurations. We take the phase space to consist of $N$ discrete
states (with label $i$), though the method applies equally to
continuous systems. Often only a tiny fraction of the phase space (the
part at low energy) is relevant to the properties being studied, due
to the strong variation of the Boltzmann weight $\exp (-\beta E_{i})$
in the canonical ensemble (CE). It is then helpful to sample in an
ensemble (with relative weights $w_{i}$ and absolute probabilities
$p_{i} = w_{i}/\sum_{j}w_{j}$) which is concentrated on this region of
phase space. The Metropolis algorithm \cite{metropolis} samples
directly in the CE, and is good at determing many physical properties
(with the notable exception of the free energy). The price to be paid
for this is that successive configurations are not independent
(typically they have a single microscopic difference), but instead
form a Markov chain with some equilibration time $t_{\rm eq}(w_{i})$.

We may distinguish two important characteristics of a Monte Carlo
simulation: its ergodicity (measured by $t_{\rm eq}(w_{i})$) and its
pertinence (measured by $N_{\rm s}(w_{i}; \cal I)$, the average number
of independent samples needed to obtain the information $\cal I$ that
we seek). We should choose $w_{i}$ so as to minimize the total number
of configurations that need to be generated, which is proportional to
$t_{\rm eq}(w_{i}) N_{\rm s}(w_{i}; \cal I)$. It is easy to specify an
ensemble which would yield the sought information if independent
samples could be drawn from it, but an ensemble with too much
weight at low energies may become fragmented into ``pools'' at the
bottoms of ``valleys'' of the energy function, and so have a large
equilibration time. For example, it is well known that at low
temperatures the Metropolis algorithm can get stuck in ordered or
glassy phases. Ergodicity may be improved by sampling instead in a
non-physical ensemble with a broad energy distribution, which allows
the valleys to be connected by paths passing through higher energies
\cite{berg2,marinariST,bergRC}. A weight assignment
leading to such a distribution cannot in general be written as an
explicit function of energy alone; rather it is an algorithm's purpose
to find this assignment, which then tells us about the density of
states $\rho(E)$. This reversal (starting with the distribution and
finding the weights) of the usual Monte Carlo process can be achieved
using a series of normal simulations, adjusting the weight $w_{i}$
after each run so that the resulting energy distribution
$\rho_{w_{i}}(E)$ converges to the desired one. Although one might
need more samples from such a broad energy ensemble (BEE) than from a
particular CE in order to find properties relating to that CE, it is
possible for a single BEE simulation to provide information on
properties over a range of temperatures. BEEs are also helpful for
finding free energies\cite{torrie2} (since relative normalizations
can only be determined for overlapping distributions
\cite{metropolis,valleauMS}) and for sampling across regions of
negative heat capacity in the vicinity of first-order phase
transitions \cite{bergMCE}.

The energy distribution used by a BEE algorithm is a free parameter
\cite{lee2}, and is often taken to be uniform (this was called the
``multicanonical ensemble'' (MCE) by Berg \cite{berg2}). It would,
however, be natural to look for an optimal most general distribution,
ie one with the best worst-case performance in terms of ergodicity and
pertinence. We will consider only monotonically decreasing weight
assignments $w_{i}$, implemented using the Metropolis scheme of
accepting a transition $i \rightarrow j$ with probability~$\mbox{\rm
min}\left(w_{j}/w_{i}, 1\right)$. Our proposal is to use the ensemble
with weight
\begin{equation} \label{eq:ONE}
w_{i} = \frac{1}{k_{i}},
\end{equation}
where $k_{i}$ is the number of states with energies up to and
including $E_{i}$. This ensemble has the property that $\log(N) N_{\rm
s}$ independent samples from it convey as much information, concerning
any property, as $N_{\rm s}$ independent samples from any rival
ensemble \cite{thing} (the factor $\log N$, which is a measure of the
ensemble's worst-case pertinence, is smaller for this ensemble than
for any other). In particular, of order $\log N$ independent samples
from this ensemble are sufficient both to find the ground state and to
determine the normalization of the density of states. While the best
worst-case ergodicity is probably obtained by sampling at infinite
temperature, this is useless in terms of pertinence. We expect
reasonable ergodicity for the $1/k$ ensemble since if we require a
rival ensemble to assign an equal probability to some state, then its
transition rates from this state may exceed those in the $1/k$
ensemble by a factor of at most $\log N$. In contrast, the
equilibration time for uniform energy sampling may be made arbitrarily
large by choosing a suitably unreasonably reparametrized hamiltonian
$H^{\prime} = f(H)$, where $f(H)$ is a monotonically increasing
function (the $1/k$ ensemble is invariant under such operations).

The $1/k$ ensemble is equivalent to uniform entropy sampling (ie
$\rho_{1/k}(E) \propto dS/dE \equiv 1/T(E)$) since for practical
purposes the entropy $S$ is given by $S(E_{i})\sim\log{k_{i}}$. Like
the CE, it has a sensible thermodynamic limit in that the relative
weight of states with a single microscopic difference remains of order
unity as $M \rightarrow \infty$, where $M$ is the system
size. However, whereas in the CE fluctuations in intensive quantities
such as energy density typically go to zero like $M^{-\frac{1}{2}}$,
in the $1/k$ ensemble they are independent of $M$, with the result
that the $1/k$ ensemble is non-self-averaging even for simple systems
such as the ferromagnetic Ising model. For example, if the physical
system has a second order phase transition at some temperature
$T_{c}$, this will be reflected by a power law contribution to the
spin-spin correlation function in the $1/k$ ensemble
\cite{cfnote}, with a new exponent:
\begin{equation} \label{eq:GONEshort}
G_{1/k}(r) - G_{1/k}(\infty) \sim
r^{-(d-2+\eta)-(1-\alpha)/\nu}.
\end{equation}
In spite of this, the correlation function (determined by spatial
average) for a state drawn from the $1/k$ ensemble is likely to be
exponentially decaying, with a random correlation legnth. To obtain
(\ref{eq:GONEshort}), we first note that uniform sampling of the
entropy leads to smooth (ie at least once differentiable) sampling of
the energy, at least in systems or regimes where the heat capacity and
temperature are strictly positive, since
\begin{equation}
\frac{d\rho_{1/k}(E)}{dE} \propto -\frac{1}{T^{2}C}.
\end{equation}
The $1/k$ ensemble may be expressed as a linear combination of canonical
ensembles (in the thermodynamic limit):
\begin{eqnarray}
p_{1/k} & \begin{array}[t]{c} \rightarrow \\[-2mm] {\scriptstyle N \rightarrow
\infty} \end{array} & \int p_{\rm CE}(T(\overline{E})) \,
\rho_{1/k}(\overline{E}) \, d\overline{E} \nonumber \\
& \equiv & \int p_{\rm CE}(T) \,
\rho_{1/k}(T) \, dT
\end{eqnarray}
where $\overline{E}$ is the normalized energy and $p$ represents any
probability assigned in an ensemble, since relative fluctuations in
the CE go to zero in the thermodynamic limit. Close to the critical
energy (letting $t=(T-T_{c})/T$) we find
\begin{equation}
\rho_{1/k}(t) \sim t^{-\alpha}
\end{equation}
where $\alpha$ is the critical exponent describing the divergence of
the heat capacity. Under a real-space renormalization with scale
factor $b$, $\rho_{1/k}(t)$ is carried by the flow ($t \rightarrow
b^{1/\nu}t$, where $\nu$ is the exponent describing the divergence of
the correlation length) away from the fixed point, and so reduces by a
factor $b^{-(1-\alpha)/\nu}$. Thus there is a contribution to
$G_{1/k}(r)$ which, as in the canonical critical ensemble, scales
under RG transformations, though with an extra factor of
$b^{-(1-\alpha)/\nu}$. This reflection of critical properties
(which normally require parameter tuning) in the $1/k$ ensemble shows that it
in some sense exhibits (by means of non-trivial probability
distributions) possible behaviours of the system over all
temperatures.

In principle, $1/k$ sampling may be implemented by an algorithm whose
only parameters \cite{transitions} are the number of Monte Carlo steps
to use at each stage of the convergence process (which should be
enough for equilibration to have occurred, and might be determined by
the algorithm). Specifically, we may represent the $n$th approximation
$\rho^{n}(E)$ to the density of states as a set of delta-functions and
use the recurrence
\begin{equation} \label{eq:stage}
\rho^{n+1}(E) = \frac{\displaystyle \sum_{\rm samples} \frac{1}{w_{i}^{n+1}}
\delta(E-E_{i})}{\displaystyle \sum_{\rm samples} \frac{1}{w_{i}^{n+1}}}
\end{equation}
with
\begin{equation} \label{eq:bias}
w_{i}^{n+1} = \left\{\begin{array}{ll} \sigma^{n}(E)^{-1} &
\hspace{4mm} \mbox{if $E \ge
E_{\rm min}^{n}$} \\ \widetilde{\sigma^{n}}(E)^{-1} & \hspace{4mm} \mbox{if $E
<
E_{\rm min}^{n}$} \end{array} \right.
\end{equation}
where $\sigma^{n}(E) = \int_{-\infty}^{E_{i}} \rho^{n}(E)\,dE$ is the
integrated density of states and $\widetilde{\sigma^{n}}(E)$ is an
extrapolation of $\sigma^{n}(E)$ below the lowest sampled energy
$E_{\rm min}^{n}$. Note that when sampling a continuous space, one
should use $w_{i}^{n+1} = (\sigma^{n}(E) + \sigma_{\rm offset})^{-1}$
in order to make the sampled entropy range finite.  $\sigma^{n}(E)$
may be evaluated in of order $\log N_{\delta}^{n}$ steps, where
$N_{\delta}^{n}$ is the number of delta-functions used to represent
$\rho^{n}$ (memory constraints may limit $N_{\delta}^{n}$ so that some
grouping procedure is needed for the delta-functions). We have also
found it useful to represent $p^{n}(E)$ as a histogram and to compute
and store the bias function $w_{i}$ before each run. One requires only
that the histogram is fine enough to resolve variation in $\rho(E)$.
Uniform energy sampling, in contrast, needs a specific choice of
histogram, which must be coarse enough to have good statistics. In
$1/k$ sampling, equation (\ref{eq:bias}) automatically interpolates as
finely as permitted by the data, short of curve fitting. However curve
fitting is helpful in determining $\widetilde{\sigma^{n}}(E)$, since
with each run the range of energies being sampled increases to cover
energies where the predicted $\sigma(E)$ used in (\ref{eq:bias}) is
not wrong by a large factor. The first run may use $w^{0}_{i} = \rm
const.$, which is likely to lead to progressively increasing
equilibration times in the following runs as the sampled energy range
extends further down.

The improved ergodicity of BEEs makes them attractive for use in hard
optimization problems \cite{marinariST,bergRC}. While their
applicability may be similar to that of simulated annealing (see eg
\cite{kirkpatrickSA}), their behaviour differs in that they offer
``open-ended'' improvement, since they never commit to a particular
valley, but continue to search for better solutions. They
also dispense with the need for a cooling schedule, which is a crucial
parameter for simulated annealing algorithms. Although a ``cautious''
BEE algorithm may spend most of its time visiting highly non-optimal
configurations, this could be offset by using parallel computation, such
as one might anticipate being readily available in the future
(equilibration time, on the other hand, is a basic constraint on an
algorithm's performance).

Our first test of $1/k$ sampling is a 100-city travelling salesperson
problem (see eg \cite{GA}), with moves consisting in segment transport
or reversal, following \cite{NR}. Fig.~1 shows $\sigma^{45}(E)$, where
each run was continued until $2\times 10^{6}$ transitions had been
accepted, and (\ref{eq:bias}) was used with the trivial extrapolation
$\widetilde{\sigma^{n}}(E<E^{n}_{\rm min}) = \sigma^{n}(E^{n}_{\rm
min})/2$.

\begin{figure}
\inseps{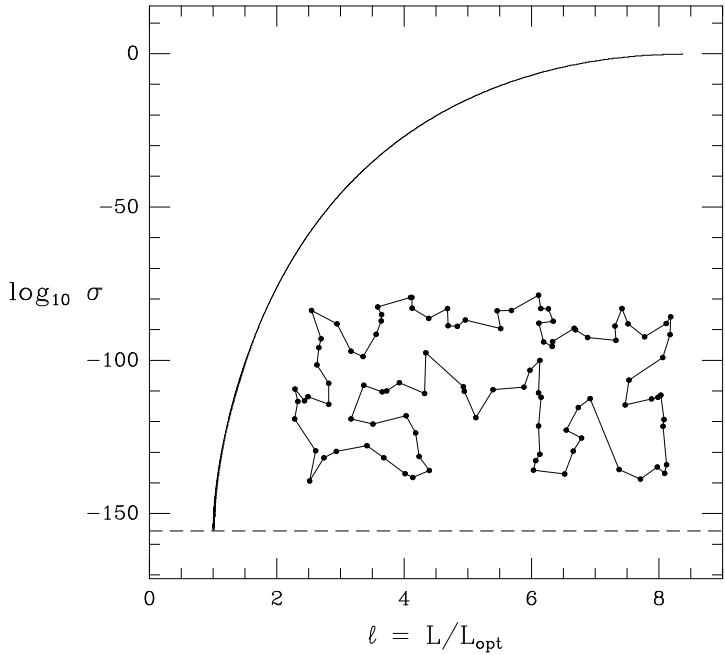}{0.5}
\vskip -30mm
\caption[a]{Integrated density of states for the archived travelling
salesperson problem ``kroA100'' \protect
\cite{reinelt,ftp}. Normalization is with respect to the established
optimal tour length $L_{\rm opt}$, as listed in the
archive. The dashed line shows $\rho_{\rm min} = 2/99!$. Inset: the
optimal tour.}
\end{figure}

\begin{figure}
\inseps{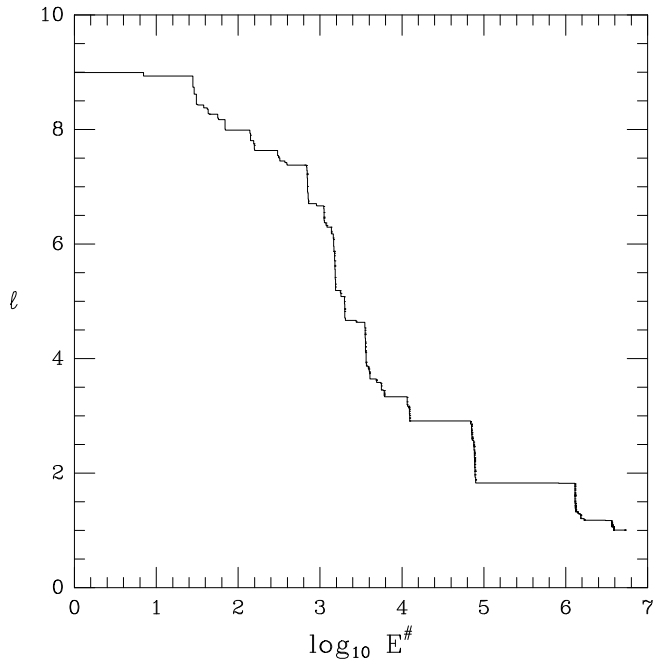}{0.5}
\vskip -30mm
\caption[a]{Length of the best-so-far tour as a function of the number of
cost evaluations $E^{\#}$, for a particular run. Among 10 such runs,
the number of cost evaluations required to find the optimal tour
varied between $\sim 2 \times 10^{6}$ and $\sim 64 \times 10^{6}$. The
plateaus are due to excursions back up to non-optimal configurations.}
\end{figure}

An additional run was conducted starting with a randomized
configuration but using the previously obtained density of states;
Fig.~2 shows the length of the best-so-far tour as a function of the
number of cost evaluations (which we consider to be more relevant than
computer time since it is more characteristic of an algorithm and
since some optimization problems, eg protein folding (which has been
studied using the MCE \cite{hansmann}), may involve expensive cost
calculations). This should be regarded as an upper limit for the
performance of $1/k$ sampling in that not all of the 45 iterations of
(\protect \ref{eq:stage}) could be eliminated by extrapolating the
density of states, and as a lower limit in that no parallelism was
used.

If we know $N$, then the absolute value of $\rho^{n}_{\rm min}$
provides a useful measure of progress during global optimization,
since $N \rho^{n}_{\rm min}$ serves as an estimate for the number of
states at or below the lowest energy sampled (assuming ergodicity). In
this way, using runs of $16 \times 10^{6}$ accepted transitions on the
problem instance shown in Fig.~1, we obtained a ground state entropy
of $0.15 \pm 0.15$~bits, with a variance of $0.6$ bits.

In order to compare $1/k$ sampling with the multicanonical ensemble,
we performed simulations of the Edwards-Anderson model with Ising
spins $s_{i}=\pm 1$ and nearest-neighbour interactions $J_{ij}=\pm 1$
(with $\Sigma J_{ij}=0$), on a $12 \times 12$ square lattice with
periodic boundary conditions.  Fig.~3 shows the energy visitation
densities $H(E)$ and the calculated entropy, $s(E) = \log 2 +
\log(\sigma(E))/12^{2}$, for one realization. For $9$ realizations we
computed the ergodicity times in sweeps (MC steps per spin), following
\cite{berg2}. We found $\tau^{\rm e}_{1/k}$ to vary between $1199$ and
$19512$, with median $2025$, while $\tau^{\rm e}_{1/k}/\tau^{\rm
e}_{\rm MCE}$ was more sharply peaked, at $0.69 \pm 0.04$. The ground
state entropies were $s(E_{0}) = 0.080 \pm 0.019$ nats per spin.

\begin{figure}
\inseps{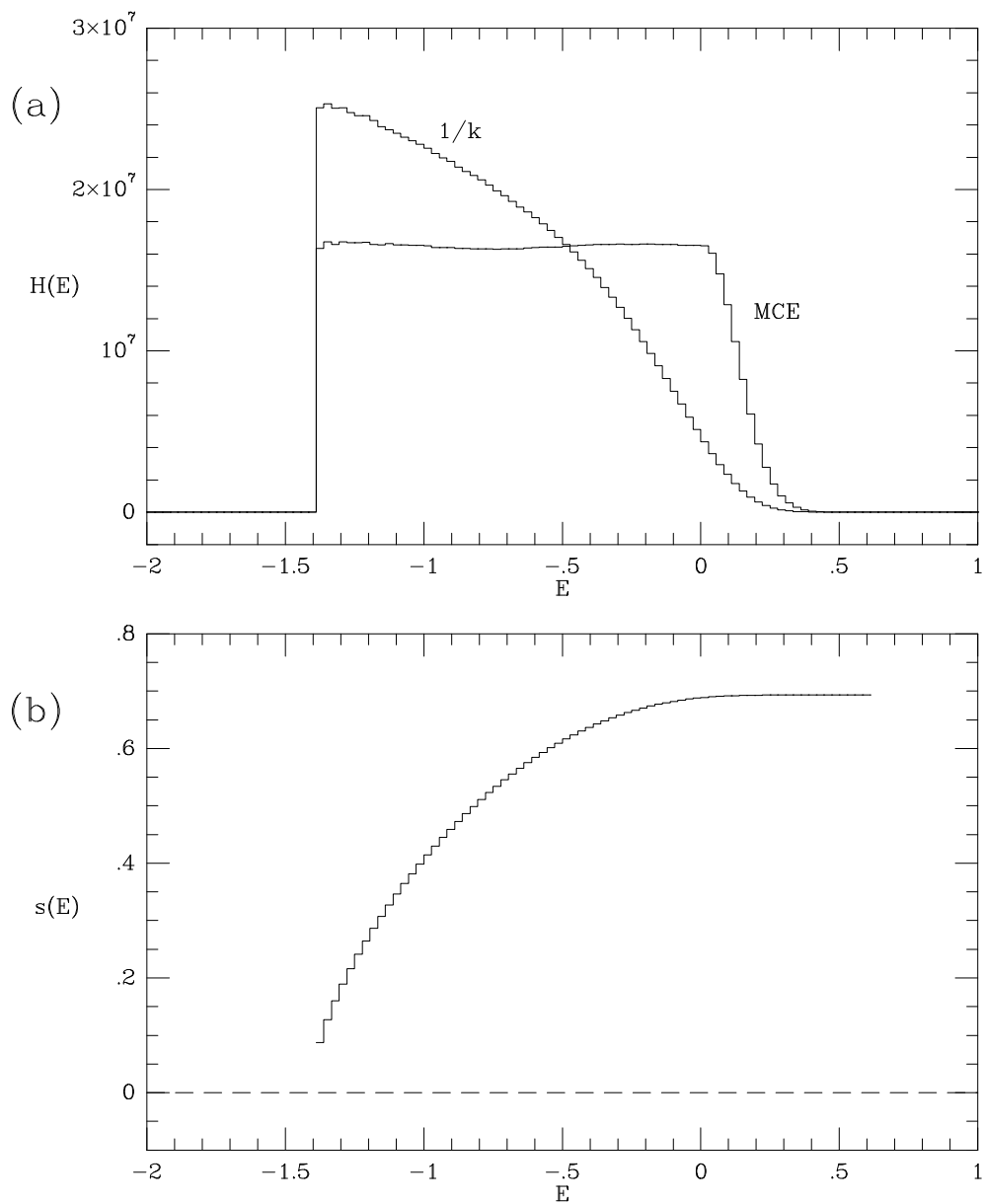}{0.7}
\caption[a]{Results for simulations using $6.4 \times 10^{6}$ sweeps on
one realization of the $12 \times 12$ Edwards-Anderson spin glass. (a)
Histogram $H(E)$ of the energy visitation density in the $1/k$
ensemble and in our implementation of the MCE following
\cite{berg2}. (b) The entropy per spin for the same system.}
\end{figure}

The last application reported here is a simulation of a regular
system with frustration, the triangular antiferromagnet, on a $48
\times 48$ parallelogram with periodic boundary conditions. Using 5
runs of $7.4 \times 10^{5}$ sweeps, we obtained a ground state entropy
of $0.32320$, with a variance of $0.00015$, which is consistent with
the exact bulk value\cite{wannier} of $(2/\pi) \int_{0}^{\pi/3} \log
(2 \cos \omega)\,d\omega \simeq 0.32307$.

These simulations show that $1/k$ sampling has significant advantages
over existing techniques. For the travelling salesperson problem it
found the global optimum, its only parameter being the number of
iterations to use. Lee and Choi \cite{lee&choi} have obtained good
results for large scale travelling salesperson problems using a
``multicanonical annealing'' algorithm which is based on the MCE, but
constrained to a certain energy range which is then ``annealed.''
While this approach is less greedy than simulated annealing, we
believe that ergodic algorithms will have a higher probability of
finding the global optimum in the limit of many samples or of much
parallelization. $1/k$ sampling may, however, benefit from being
truncated above some fixed energy, provided this isn't so low as to
compromise ergodicity. The results for the spin glass show that $1/k$
sampling has faster equilibration and more weight for low-lying states
than the MCE, though it would be worthwhile to continue this
comparison to larger systems. It would also be interesting to compare
the variance of the ground state entropy results for the triangular
antiferromagnet with that obtained by other methods.

$1/k$ sampling may also be useful for determining the functional form
of a density of states, since it is completely impartial on account of
its reparametrization invariance. Unfortunately the equilibration
times of BEE algorithms are rather long, going as $M^{2}$ in the best
case and as more than $M^{3}$ for the Edwards-Anderson spin glass
\cite{berg2}. While BEE algorithms may be unnecessarily cautious for
studying well-behaved systems when free energies are not required, the
slower equilibration for the spin glass probably reflects the
intrinsic difficulty of this problem. We suggest that $1/k$ sampling
may be especially useful for obtaining complete and reliable
information on the properties of relatively small systems, since it
has, among a large class of ensembles, the most general applicability
in terms of the number of independent samples needed, combined with
robust ergodicity properties and a minimum requirement for input from
the user.


\begin{thebibliography}{10}
\bibitem{thing}
This is because $\log(N) N_{\rm s}$ samples from the $1/n$ ensemble may be
  processed (by deleting each sample with a suitable probability) to produce
  $N_{s}$ samples from a rival ensemble $w^{\prime}_{i}$. The $1/n$ ensemble is
  optimal in this sense since we could want to know about properties
  arbitrarily close to some energy $E_{i}$, in which case
  $(p^{\prime}_{i}/p_{i})N_{\rm s}$ samples from $w_{i}$ tell us the same as
  $N_{\rm s}$ samples from $w^{\prime}_{i}$, and the probability assigned by
  the $1/n$ ensemble is everywhere smaller by the same factor (viz $\log{N}$)
  in comparison with the greatest probability that any decreasing distribution
  could assign.
\bibitem{cfnote}
Note that the connected part of the correlation function $G_{1/n}(r)$ does not
  tend to zero at infinity since no symmetry is broken in the $1/n$ ensemble -
  distant spins really are correlated. A function which does go to zero is
  $G_{1/n}^{\prime} = \left<m_{i}m_{j}\right> - \left<m_{i}\,\mbox{\rm
  sign}(S)\right>\left<m_{j}\,\mbox{\rm sign}(S)\right>$, in which the
  subtracted part may be interpreted as the value of $\left<m_{i}m_{j}\right>$
  that would be predicted from separate observation of each spin together with
  the total spin $S$.
\bibitem{transitions}
We regard the choice of allowed transitions as being part of the problem
  specification since it is problem-specific. A good choice, which should as
  far as possible allow progressive energy reductions, is nevertheless very
  important.
\bibitem{ftp}
The TSP library may be obtained by anonymous ftp to titan.cs.rice.edu as the
  file ``tsplib.tar'' in the directory ``public''.
\end{thebibliography}
\end{document}